\documentclass[a4paper,12pt]{article}
\usepackage{amssymb}
\newcommand\bq{\begin{equation}}
\newcommand\eq{\end{equation}}
\newcommand{\bqn}{\begin{eqnarray}}
\newcommand{\eqn}{\end{eqnarray}}
\newcommand{\nb}{\nonumber}

\begin{document}
\title{Rigidly rotating dust in general relativity}
\author{Jos\'e C.N. de Araujo\thanks{Divis\~ao~de~astrof\'\i sica,~
Instituto~Nacional~de~Pesquisas~Espaciais, \newline Avenida~dos~Astronautas,
1758, S\~ao Jos\'e dos Campos, S.P. 12227-010 , Brazil, \newline
e-mail:~jcarlos@das.inpe.br} \\
\small and \\
Anzhong Wang\thanks{Departamento de F\' {\i}sica Te\' orica, Universidade do
Estado do Rio de Janeiro, \newline  Rua S\~ ao Francisco Xavier 524, Rio de
Janeiro, R.J. 20550-013, Brazil, \newline
e-mail: wang@dft.if.uerj.br}}
\maketitle
\vspace{24pt}
\begin{abstract}
A solution to the Einstein field equations that represents a
rigidly rotating dust accompanied by a thin matter shell of the
same type is found.
\vspace{12pt}
\end{abstract}
\newpage
\section{Introduction}

Recently, Bonnor studied axially symmetric stationary solutions of
Einstein field equations coupled with dust and showed the reasons
why a density gradient parallel to the axis is allowed for General
Relativity but not for Newtonian mechanics \cite{Bonnor1977}. He
also considered an analytic solution which represents a rigidly
rotating dust, and found that the solution is asymptotically flat
in all the three spatial directions but with a total mass equal to
zero. He attributed this to the center singularity that was
believed to have an infinitely large negative mass, which just
balances the positive infinite mass of the dust that fills all the
spacetime. For the details, we refer the readers to
\cite{Bonnor1977}. Motivated by this particular solution, Bonnor
wondered whether or not a non-singular rotating dust exists.  In
the present article, we obtain a solution that generates an
axially symmetric rigidly rotating dust, which is accompanied by a
rigidly rotating thin disk, namely, a singular hypersurface
perpendicular to the axis of the rigidly rotating dust.

It is worth mentioning that there is a paper by Georgiou
\cite{Georgiou1994}, who studied rotating Einstein-Maxwell
fields\footnote{It has been one of the referees who brought to our
attention this paper by Georgiou.}. Georgiou obtains exact
exterior and matching interior stationary axially symmetric
solutions for a rigidly rotating charged dust. His solution
generates an infinitely long cylinder and a thin singular disk
perpendicular to the axis of the cylinder. Later on we consider
the differences and the similarities of our solution and that of
Georgiou's.

The present paper has also been motivated by a study by Opher,
Santos and Wang (OSW) \cite{osw1996} concerning the origin of
extragalactic jets. These authors argued that, under certain
circumstances, the spacetime given by what they refer to as the
van Stockum metric, which is associated with a dust cylinder
(studied by van Stockum \cite{vanStockum1937} and extensively
analyzed by Bonnor \cite{Bonnor1980,Bonnor1992}), can account for
the collimating effect present in extragalactic jets. OSW showed
that this dust cylinder produces confinement for the geodesic
motion of test particles for certain values of the radial energy
and angular momentum. In fact, it was one of our aims to improve
the OSW's model by looking for a spacetime that could more
realistically describe a jet.

It is worth mentioning, however, that van Stockum in fact
rediscovered a solution which was first obtained by Lanczos
\cite{Lanczos1924}\footnote{The referees brought to our attention
that Lanczos' paper preceded that by van Stockum.} . Hereafter,
therefore, we refer to the metric related to the dust cylinder as
Lanczos metric.

The extragalactic jets are ubiquitous in active galaxies, they are
highly collimated and the matter which forms them is highly
relativistic \cite{Bland1993}. It is worth mentioning that there
is no consensus in the literature to explain why they are the way
they are. Many authors assume that the jets propagate along a
direction provided by, most probably, rapidly rotating Kerr black
holes present at the centers of active galaxies (see, e.g.,
Begelman, Blandford and Rees \cite{bbr1984} and also
\cite{rees97,luminet98}). This fact also suggest that putative
general relativistic effects could be important (see, e.g.,
\cite{bsh1993,fc1997,rk1996,rk1997,ksk98}, among others).

The central engine that gives rise to the jets could be more
complex than a simple super massive black hole, it could well
occur that jets be driven by an axially symmetric structure
present at the center of active galaxies. The Lanczos solution
(referred to by OSW as van Stockum solution), however, has its
weakness when applied to an actual physical situation: it
represents an infinitely long cylinder. As a result, the spacetime
is not asymptotically flat and has infinite mass.

The solution here studied generates an axially symmetric rigidly
rotating dust accompanied by a surface layer, which does not
satisfy the energy conditions (i.e., weak, dominant and strong
\cite{he1973}) in part of the hypersurface. We point out however,
that other authors (see Refs. \cite{Taub1980,sws1997,wss1997}, and
references cited therein) have investigated such structures and
some of them, as in our case, do not satisfy any of the energy
conditions. In some of these cases, the energy conditions may be
satisfied by a suitable choice of parameters.

We argue that the present study may be by itself of interest,
because it represents a new axially symmetric dust solution, which
could also motivate other authors to find solutions physically
satisfactory.

In section 2 a closed form of the solutions are given and the main
properties of them are studied, while in section 3 our main
conclusions and some discussions are presented.

\section{The Rotating Dust Metric}

Our starting point is the Lanczos metric given by
\cite{Lanczos1924} \bq ds^2 = dt^2 - 2k dt d\varphi - l d\varphi^2
- e^{\mu}(dr^2 + dz^2), \eq where \bqn k &=& \alpha\eta\;, \;\;\;
\eta = r\xi_{,r}\;,
 \;\;\; l = r^2 - \alpha^2\eta^2\;, \nb\\
\mu_{,r} &=& \frac{\alpha^{2}}{2r} (\eta^2_{,z} -
\eta^2_{,r})\;,\;\;\;\; \mu_{,z}= -
\frac{\alpha^{2}}{2r}\eta_{,r}\eta_{,z}\;, \eqn  and the function
$\xi(r,z)$ satisfies the Laplacian equation
$\bigtriangledown^{2}\xi = 0$, with $\bigtriangledown^{2}$ being
the Laplacian operator in Euclidean three-space. The symbol $(
)_{,x}$ denotes partial derivative with respect to the argument
$x$, and $\{x^{\mu}\} \equiv \{t, r, z, \varphi\},\; (\mu = 0, 1,
2, 3)$ are the usual axisymmetric coordinates. One can show that
the above solutions satisfy the Einstein field
equations\footnote{In this paper we choose units such that $G = 1
= c$, where $G$ is the gravitational constant, and $c$ the speed
of light.} $R_{\mu\nu} - g_{\mu\nu}R/2 = - 8\pi \rho
u_{\mu}u_{\nu}$ with the energy density and four-velocity of the
dust being given, respectively, by \bq \rho=\frac{e^{-\mu}}{8\pi
r^{2}}(\eta^2_{,z}+\eta^2_{,r}), \;\;\; u^{\mu} =
\delta^{\mu}_{0}. \eq The above solutions represent rigidly
rotating dust. This can be seen, for example, by calculating the
shear in this non-expanding spacetime, $q_{\mu\nu} \equiv
(u_{\mu;\nu} + u_{\nu;\mu})/2$, which is identically zero for the
solutions given by Eqs.(1) and (2). However, the angular velocity
of the dust, which is given by $w_{\mu\nu} \equiv (u_{\mu;\nu} -
u_{\nu;\mu})/2$, does not vanish.

The specific solution considered by Bonnor \cite{Bonnor1977}, for
example, is $\xi = 2h/\sqrt{r^{2} + z^{2}}$ with $h$ being a
constant. As shown in  \cite{Bonnor1977} this spacetime is free of
any spacetime singularities, except for that located at the origin
of the coordinate system, namely, $r = z = 0$. This singularity is
a curvature singularity with an infinitely large negative mass.

In this paper, we consider the solution with $\xi  = J_{0}(r)e^{- z}$,
where $J_{0}(r)$ denotes the zero-order Bessel function. Then,
substituting it into Eqs.(2) and (3) we find
\bqn
k &=& \alpha r J_{1}(r) e^{- z}, \;\;\;
l = r^{2}\left[1 - \alpha^{2}J_{1}^{2}(r)
e^{- 2z}\right], \nb\\
\mu & = & - \frac{\alpha^{2} r}{2} J_{0}(r) J_{1}(r) e^{- 2z}, \\
\rho & = & \frac{\alpha^{2}e^{ -2z}}{8 \pi}\left[J_{0}^{2}(r) +
J_{1}^{2}(r)\right]e^{- \mu}.
\eqn

From the above equations we can see that the spacetime is singular
when $z \rightarrow - \infty$. To remedy this undesirable feature
we can replace $z$ by $|z|$ in Eq.(4), i.e., \bqn k &=& \alpha r
J_{1}(r) e^{- |z|}, \;\;\; l = r^{2}\left[1 -
\alpha^{2}J_{1}^{2}(r) e^{- 2|z|}\right], \nb\\ \mu & = & -
\frac{\alpha^{2} r}{2} J_{0}(r) J_{1}(r) e^{- 2|z|}. \eqn

Before proceeding it is worth mentioning  that in a paper by
Georgiou \cite{Georgiou1994} exact exterior and matching interior
solutions are found, where the interior solution is similar to the
solution present here. As shown in \cite{Georgiou1994} the
spacetime refers to solutions of the Einstein-Maxwell field
equations for a rigidly rotating charged dust with vanishing
Lorentz force. The solutions generate an infinitely long cylinder
of charged dust rigidly rotating about its axis and a 4-current
located on a singular hypersurface perpendicular to the axis of
the cylinder at the origin of the coordinate system.

Due to the fact that Georgiou's interior solutions present a non
null electromagnetic 4-potencial, namely, $A_{\mu}=(0,0,0,A_{3})$,
his equations related to the $\mu$ function involves the $A_{3}$
function (see Eqs.(4.2) and (4.3) in \cite{Georgiou1994}). In such
a way he could set $F=e^{\mu}=1$. On the other hand, we have $F=1$
and $\mu$ is given by Eq.(4), as a result our solutions are
different. Apart from the constants our mass density contains in
addition the term $e^{-\mu}$, as a result a stronger dependence on
the z coordinate occurs as compared to the Georgiou's mass
density. Our solution describes rigidly rotating neutral dust and
Georgiou's solution, rigidly rotating charged dust. Consequently
the resulting spacetimes are different.

One can show that such resulted spacetime present here is
asymptotically flat in the z direction, since as $|z|\rightarrow
\infty$ for any particular finite value of $r$, $k \rightarrow 0$,
$l \rightarrow r^{2}$ and $e^{\mu} \rightarrow 1$. On the other
hand, the behaviour of the solution as $r \rightarrow \infty$, for
any particular finite value of z, shows that, for example, $k$
oscillates infinitely between $-\infty$ and  $+\infty$. This would
indicate that the spacetime is not asymptotically flat.

On the other hand, the Kretschmann scalar is given by \bqn
{\cal{R}} &\equiv&
R_{\alpha\beta\gamma\delta}R^{\alpha\beta\gamma\delta} = -
\frac{\alpha^{2}e^{- 2(|z| + \mu)}}{4r^{2}}\left\{
16\left[J_{1}^{2}(r) - rJ_{0}(r)J_{1}(r)\right.\right.\nb\\ &&
\left. + r^{2}\left(J_{0}^{2}(r) + J_{1}^{2}(r)\right) \right] -
4\alpha^{2}r^{2}e^{- 2|z|} \left[2J_{0}^{4}(r) +
J_{1}^{4}(r)\right.\nb\\ && \left. + 7J_{0}^{2}(r)J_{1}^{2}(r) -
4rJ_{0}(r)J_{1}(r) \left(J_{0}^{2}(r) +
J_{1}^{2}(r)\right)\right]\nb\\ && \left. + \alpha^{4}r^{4}e^{-
4|z|}\left[J_{0}^{6}(r) + J_{1}^{6}(r) +
3J_{0}^{2}(r)J_{1}^{2}(r)\left(J_{0}^{2}(r) + J_{1}^{2}(r)\right)
\right]\right\} \nb\\ && + {\cal{R}}_{0}\delta(z), \eqn where
${\cal{R}}_{0}(r)$ is a bounded function of $r$ (see the
discussions following Eq.(14) below), and $\delta(z)$ denotes the
Dirac delta function. Using the relations \bqn J_{n}(x) \approx
\left\{\begin{array}{ll} \frac{x^{n}}{2^{n}n!}, & x \rightarrow
0,\\
\\
\sqrt{\frac{2}{\pi x}}\cos\left(x - \frac{2n + 1}{4}\pi\right), &
x \rightarrow + \infty,\\ \end{array}\right . \eqn we can see from
Eq.(7) that ${\cal{R}} \rightarrow $ finite, as $r \rightarrow 0$,
and that ${\cal{R}} \rightarrow 0$, as $|z|$ or $r \rightarrow +
\infty$ indicating that the spacetime is asymptotically flat.
Although we have considered here some discussion on flatness, it
is worth bearing in mind that some authors argue that this concept
is not well defined (see,e.g., \cite{Stewart1993}).

Also, from Eqs.(6) and (8) we have \bqn X \equiv
||\partial_{\varphi}||^{2} = |g_{\varphi\varphi}| &\rightarrow&
O(r^{2}), \nb\\ \frac{X_{,\alpha}X^{,\alpha}}{4X} &\rightarrow& 1,
\eqn as $r \rightarrow 0$. Hence, the axis ($r = 0$) of the
spacetime is well defined and  locally flat. From the above
equations we can also see that, by properly choosing the constant
$\alpha$, we may have $g_{\varphi\varphi} < 0$ for any $r$. That
is, the spacetime may be free of any closed time-like curves.
Therefore, the solution given by Eq.(6) represents an axially
symmetric and rigidly rotating dust spacetime.

It should be noted that the replacement of $z$ by $|z|$ gives rise
to a thin matter shell. As a matter of fact,  this replacement
mathematically is equivalent first to cut the original spacetime
given by Eq.(4) into two parts, $z > 0$ and $z < 0$, and then join
the part $z > 0$ with a copy of it along the hypersurface $z = 0$,
so that the resulted spacetime has a reflection symmetry with
respect to the surface. After this cut-paste operation, the
spacetime is no longer analytic across the surface $z = 0$.
Actually, the metric coefficients are continuous, but their first
derivatives with respect to $z$ are not. Then, according to Taub's
theory \cite{Taub1980,Note}, a thin matter shell appears on the
hypersurface. Introducing the quantity $b_{\mu\nu}$ via the
relation \bq b_{\mu\nu} \equiv \left. g^{+}_{\mu\nu,z}\right|_{z =
0^{+}} - \left. g^{-}_{\mu\nu,z}\right|_{z = 0^{-}}, \eq where
$g^{+}_{\mu\nu} \;(g^{-}_{\mu\nu})$ are quantities defined in the
region $z > 0 \;(z < 0)$, we find that the non-vanishing
components of $b_{\mu\nu}$ are given by \bqn b_{11} &=& b_{22} = -
2 \alpha^{2} r J_{0}(r) J_{1}(r) e^{\mu_{0}},\nb\\ b_{03} &=&
b_{30} = 2 \alpha r J_{1}(r), \;\;\; b_{33} = - 4 \alpha^{2} r^{2}
J_{1}^{2}(r), \eqn where $\mu_{0} \equiv \mu(r,z)|_{z = 0}$. Then,
the surface energy-momentum tensor $\tau_{\mu\nu}$ is given by
\cite{Taub1980} \bq \tau_{\mu\nu} = \frac{1}{16\pi}\left\{ b(n
g_{\mu\nu} - n_{\mu}n_{\nu}) +
n_{\lambda}(n_{\mu}b^{\lambda}_{\nu} + n_{\nu}b^{\lambda}_{\mu}) -
(n b_{\mu\nu} + n_{\lambda}n_{\delta} b^{\lambda
\delta}g_{\mu\nu})\right\}, \eq where $n_{\mu}$ is the normal
vector to the hypersurface $z = 0$, given by $n_{\mu} =
\delta^{2}_{\mu}$, with $n \equiv n_{\lambda}n^{\lambda}$ and $b
\equiv b_{\lambda}^{\lambda}$. Substituting Eq.(11) into Eq.(12),
we find that $\tau_{\mu\nu}$ can be written in the form \bq
8\pi\tau_{\mu\nu} = \sigma u_{\mu}u_{\nu} + p x_{\mu}x_{\nu} + q
(u_{\mu}x_{\nu} + u_{\nu}x_{\mu}), \eq with \bqn \sigma &=& - p =
- \alpha^{2} r J_{0}(r)J_{1}(r)e^{- \mu_{0}},\nb\\ q &=& \alpha
J_{1}(r) e^{- \mu_{0}}, \eqn where $u_{\mu}$ is the four-velocity
of the dust restricted to the surface $z = 0$, and $x_{\mu}$ is a
space-like unit vector on the surface, given by $x_{\mu} = r
\delta^{3}_{\mu}$, and has the properties: $x_{\lambda}x^{\lambda}
= - 1$ and $x_{\lambda}u^{\lambda} = 0$. The non null components
of $\tau^{\mu}_{\;\;\nu}$ are explicit shown in the appendix,
where we also obtain them using the alternative technique derived
by Israel \cite{Israel1966}.

Eq.(13) shows that $\sigma$ represents the surface energy density
of the shell measured by observers who are comoving with the dust
fluid, and $p$ is the pressure in the direction perpendicular to
the observers' world lines, while $q$ represents the heat flow. On
the other hand, Eqs.(14) show that all these quantities are finite
for any $r$. One can also see that $\sigma$ oscillates between
positive and negative values; when $\sigma < 0$ the energy
conditions \cite{he1973} are violated. Note that the function
${\cal{R}}_{0}(r)$ appearing in Eq.(7) is a combination of them
and is finite, too.

Note that the quantities $q_{\mu\nu}$ and $w_{\mu\nu}$ defined
above contain only first derivatives of the metric coefficients,
as a result they can be written generally in the form $$
Y_{\mu\nu} = Y^{+}_{\mu\nu}H(z) + Y^{-}_{\mu\nu}H(- z), $$ where
$Y^{\pm}_{\mu\nu} (\equiv \{q^{\pm}_{\mu\nu}, w^{\pm}_{\mu\nu}\})$
are quantities calculated respectively in the regions $z > 0$ and
$z < 0$, and $H(z)$ is the Heavside function, which is $1$ for $z
> 0$, $1/2$ for $z = 0$, and $0$ for $z < 0$.  Since
$q^{\pm}_{\mu\nu}$ are all equal zero, we can see that
$q_{\mu\nu}$ is zero even on the surface $z = 0$. That is, the
matter shell is also rigidly rotating.

On the other hand, in terms of $\tau_{\mu\nu}$, the Einstein field
equations for the solution (6) takes the form \bqn R_{\mu\nu} -
\frac{1}{2} g_{\mu\nu}R = - 8\pi [\rho u_{\mu}u_{\nu} +
\tau_{\mu\nu}\delta(z)], \eqn where $\rho$ is given by Eq.(5) but
with $z$ being replaced by $|z|$. Thus, the solution given by
Eq.(6) represents a rigidly rotating dust accompanied by a matter
shell of the same type.

To further study the spacetime of this solution, let us consider
the total mass of the spacetime. Using the formula of Tolman
\cite{Tolman1987}, we find that the mass inside the cylinder $r =
R$ is given by \bqn M(R) &=& \int^{2\pi}_{0}\int^{+ \infty}_{-
\infty}\int^{R}_{0} \left\{t^{0}_{\;0} + \sqrt{- g}\left[\rho
u^{0}u_{0} + \tau^{0}_{\;\;0}\delta(z)\right]\right\}d\varphi dzdr
\nb\\ &=& \frac{\alpha^{2} R}{16}\left\{2 J_{0}(R)J_{1}(R) +
R\left[J_{0}^{2}(R) - J_{1}^{2}(R)\right]\right\}, \eqn where $R$
is a constant, and $t_{\;\mu}^{\nu}$ is the so-called
gravitational energy-momentum pseudo-tensor \cite{Tolman1987}. The
combination of Eq.(8) and Eq.(16) shows that $M(R)$ oscillates
infinitely between $- \infty $ and $+ \infty $ as $R \rightarrow +
\infty $. Thus, in the present case the total mass of the
spacetime is not well defined. Were $\sigma$ always negative it
could occur that the total mass of the dust structure were zero.
This behaviour of $M(R)$ as $R \rightarrow \infty$ is related to
the fact that $\sigma$ oscillates between positive and negative
values. To remedy this problem, one might cut the spacetime and
smoothly match the dust structure where $\sigma
>0$  to vacuum spacetimes. As result one would have a hollow
rigidly rotating dust structure with two vacuums, one inside and
other outside it. It could occur however that this procedure
generated unphysical surface layers. Due to the complexity of
these issues, we have not yet been successful in this direction.

\section{Discussion and Conclusions}

Motivated by the article by Bonnor \cite{Bonnor1977} we have
studied a particular axially symmetric rigidly rotating dust
solution, and found that it is accompanied by a thin disk located
on an hypersurface perpendicular to the symmetry axis.  The
undesirable feature is that the thin disk has negative energy
density in part of the hypersurface.

The solution we have found is in some sense similar to a solution
found by Georgiou \cite{Georgiou1994}. This author obtains exact
interior and matching exterior axially symmetric solutions of the
Einstein-Maxwell fields equations, for rigidly rotating charged
dust. The fact that our metric function $\mu$ be given by Eq.(4),
instead of being $\mu=1$ as in \cite{Georgiou1994}, turns the
spacetimes different. Therefore, the fact that the dust be charged
modify significantly the spacetime generated.

A completely satisfactory solution for  rigidly rotating dust
fluids is yet to be derived and deserves to be investigated. Such
a solution must be asymptotically flat, have finite mass, have
non-singularities, and whether a thin shell appears it must
satisfy the energy conditions.

To see whether or not our solution  allow confinement, one needs
to follow a similar procedure as in \cite{osw1996}. However, due
to the fact that the metric functions now depend on both of the
$r$ and $z$ coordinates, the study of geodesic motions become very
complicated. We therefore leave such an issue for a future study.

\vspace{36pt}

\par\noindent{\bf Acknowledgments}

We would like to thank Drs R. Opher and N.O. Santos for helpful
discussions, and O.D. Miranda for reading the paper. This work was
partially developed while J.C.N.A. had a post-doc position at
Departamento de Astronomia - Instituto Astron\^omico e Geof\'\i
sico (Universidade de S\~ao Paulo). The financial assistance from
CNPq and FAPESP is gratefully acknowledged. Finally, we would like
to thank the referees for the suggestions and criticisms that
helped us to improve our paper.

\newpage
\appendix
\section{Alternative Calculation of $\tau_{\mu\nu}$}

Here we show that Taub's  and  Israel's techniques to calculate
the surface energy-momentum tensor, $\tau_{\mu\nu}$, are
equivalent.

We start from

\bqn 8\pi\tau^{\mu}_{\;\;\nu} & \equiv & 8\pi
\lim_{\varepsilon\rightarrow 0}\;
\int_{-\varepsilon}^{+\varepsilon} T^{\mu}_{\;\;\nu}\; dz\; = -
\lim_{\varepsilon\rightarrow 0}\;
\int_{-\varepsilon}^{+\varepsilon}\left( R^{\;\mu}_{\;\;\nu}-
\frac{1}{2}\;\delta^{\mu}_{\;\nu}\;R\right)\; dz \nb \\ &=& -
\lim_{\varepsilon\rightarrow 0}\;
\int_{-\varepsilon}^{+\varepsilon}\left(
\mathfrak{R}^{\;\mu}_{\;\;\nu}-
\frac{1}{2}\;\delta^{\mu}_{\;\nu}\;\mathfrak{R}\right)\; dz
 \eqn

\par\noindent (see, e.g., \cite{Georgiou1994,Israel1966}).
\bigskip

Due to the presence of the thin disk, $R^{\;\mu}_{\;\;\nu}$ and
$R$ contain the terms $\mathfrak{R}^{\mu}_{\;\;\nu}$ and
$\mathfrak{R}$, respectively. These terms contain delta functions,
and  appear only with components of $R^{\;\mu}_{\;\;\nu}$ whose
expressions contain second order partial derivatives with respect
to z. Following \cite{Georgiou1994}, it is easy to show that the
non null components of $\mathfrak{R}^{\mu}_{\;\;\nu}$ read

\bqn  \mathfrak{R}^{1}_{\;\;1} &=& \mathfrak{R}^{2}_{\;\;2} =
{1\over 2}\;\;e^{-\mu}\;\;(\mu^{-}_{,z}-\mu^{+}_{,z})\;\;\delta(z)
\nb \\ - \mathfrak{R}^{3}_{\;\;3} &=& \mathfrak{R}^{0}_{\;\;0} =
{1\over 2}\;\;e^{-\mu}\;\;{k\over r^{2}}
\;\;(k^{-}_{,z}-k^{+}_{,z}) \;\;\delta(z) \nb \\
\mathfrak{R}^{3}_{\;\;0} &=& {\;\;e^{-\mu}\over
2r^{2}}\;\;(k^{-}_{,z}-k^{+}_{,z}) \;\;\delta(z) \nb \\
\mathfrak{R}^{0}_{\;\;3} &=& 2\;k\;\mathfrak{R}^{3}_{\;\;3} -
l\;\;\mathfrak{R}^{3}_{\;\;0}\;, \eqn

\par\noindent where the functions with superscript $+$ ($-$) stand for
the functions defined in the region $z > 0 \;(z < 0)$.

Substituting Eq.(18) into (17)  it is straightforward to show that
the non null components of $\tau^{\mu}_{\;\;\nu}$ are given by:

\bqn 8\pi\tau^{0}_{\;\;0} &=& -\alpha^{2}[ J_{1}^{2}(r)+ r
J_{0}(r)J_{1}(r) ]e^{-\mu_{0}} \nb \\ 8\pi\tau^{0}_{\;\;3} &=&
\alpha r J_{1}(r) [ 1+ \alpha^{2} J_{1}^{2}(r)] e^{-\mu_{0}} \nb
\\ 8\pi\tau^{3}_{\;\;0} &=& - {\alpha \over r} J_{1}(r)
e^{-\mu_{0}} \nb
\\ 8\pi\tau^{3}_{\;\;3} &=& \alpha^{2}[ J_{1}^{2}(r)- r J_{0}(r)J_{1}(r)
]e^{-\mu_{0}}\;,\eqn

\par\noindent which agree with Eqs.(13), showing, therefore,
that Taub's and Israel's techniques are equivalent.

Finally, it is worth mentioning that Georgiou's definition for
$\tau^{\mu}_{\;\;\nu}$ is a little bit different from Israel's.
The former integrates $T^{\mu}_{\;\;\nu}$ with respect to the
proper distance measured perpendicularly through the thin disk. On
the other hand, Eq.(17) is an integral of $T^{\mu}_{\;\;\nu}$ with
respect to the $z$ coordinate. If one followed Georgiou's
definition, instead of having the term $e^{-\mu_{0}}$ in Eq.(19)
one would have $e^{-\mu_{0}/2}$.


\begin{thebibliography}{99}
\bibitem{Note} It should be noted that the
Taub method is equivalent to that of Israel \cite{Israel1966} when
the hypersurface is non-null. As a matter of fact, introducing the
extrinsic curvature of the hypersurface by$ K^{\pm}_{\mu\nu} =
(n_{\alpha;\beta} - n_{\delta} \Gamma^{\pm \delta}_{\alpha\beta})
p^{\alpha}_{\mu}p^{\beta}_{\nu}, $ where $p^{\alpha}_{\mu}$ is the
first fundamental form of the hypersurface, defined as $
p_{\mu\nu} = g_{\mu\nu} - n_{\mu} n_{\nu}/n_{\alpha}n^{\alpha}$,
the Einstein field equations on the hypersurface take the form
\cite{Taub1980} $[K_{\mu\nu}]^{-} - p_{\mu\nu}[K]^{-} = 8\pi
\tau_{\mu\nu}$, which are exactly the ones found by Israel
\cite{Israel1966}, where $[K_{\mu\nu}]^{-} \equiv K^{+}_{\mu\nu} -
K^{-}_{\mu\nu}$, and $[K]^{-} \equiv [K_{\mu\nu}]^{-}g^{\mu\nu}$.
In the appendix we show explicitly that both techniques are
equivalent.
\end{thebibliography}
\end{document}